\documentclass{iau}
\usepackage{graphicx}

\title[Chemical peculiarities in HAeBe stars] 
{Chemical peculiarities in magnetic and non-magnetic pre-main sequence A and B stars}

\author[Folsom et al. ]   
{C.P. Folsom$^{1,2}$, S. Bagnulo$^{2}$, G.A. Wade$^{3}$, J.D. Landstreet$^{2,4}$, \\ E. Alecian$^{5}$}

\affiliation{
$^{1}$Institut de Recherche en Astrophysique et Plan\'etologie, Toulouse, France\\
email: {\tt colin.folsom@irap.omp.eu} \\
$^{2}$Armagh Observatory, Armagh, Northern Ireland\\
$^{3}$Department of Phyics, Royal Military College of Canada, Kingston, Canada\\
$^{4}$Physics \& Astronomy Department, University of Western Ontario, London, Canada\\
$^{5}$Observatoire de Paris, Meudon, France\\
}

\pubyear{2013}
\volume{302}  
\pagerange{1--2}
\jname{Magnetic fields throughout stellar evolution}
\editors{P. Petit ... eds.}
\begin{document}

\maketitle

\begin{abstract}
In A- and late B-type stars, strong magnetic fields are always associated with Ap and Bp chemical peculiarities.  However, it is not clear at what point in a star's evolution those peculiarities develop.  Strong magnetic fields have been observed in pre-main sequence A and B stars (Herbig Ae and Be stars), and these objects have been proposed to be the progenitors of Ap and Bp stars.  However, the photospheric chemical abundances of these magnetic Herbig stars have not been studied carefully, and furthermore the chemical abundances of 'normal' non-magnetic Herbig stars remain poorly characterized.  To investigate this issue, we have studied the photospheric compositions of 23 Herbig stars, four of which have confirmed magnetic fields.  Surprisingly, we found that half the non-magnetic stars in our sample show $\lambda$ Bootis chemical peculiarities to varying degrees.  For the stars with detected magnetic fields, we find one chemically normal star, one star with $\lambda$ Boo peculiarities, one star displaying weak Ap/Bp peculiarities, and one somewhat more evolved star with somewhat stronger Ap/Bp peculiarities.  These results suggests that Ap/Bp peculiarities are preceded by magnetic fields, and that these peculiarities develop over the pre-main sequence lives of A and B stars.  The incidence of $\lambda$ Boo stars we find is much higher than that seen on the main sequence.  We argue that a selective accretion model for the formation of $\lambda$ Boo peculiarities is a natural explanation for this remarkably large incidence. 

\end{abstract}

\firstsection 
\section{Introduction}

Recently, strong magnetic fields have been found in some Herbig Ae and Be (HAeBe) stars.  These are pre-main sequence A- and B-type stars ($\sim$2 to $\sim$10 $M_{\odot}$).  Magnetic fields have been found in 5-10\% of HAeBe stars: 3/50 stars by \cite{Wade2007}, and 5/70 stars by \cite{Alecian2013} in currently the most comprehensive and accurate study.  The magnetic fields found in these stars are geometrically simple, predominately dipolar with dipole strengths on the order of 1 kG (e.g. \cite[Alecian et al. 2008]{Alecian2008}, \cite[Folsom et al. 2008]{Folsom2008}).  

The incidence and morphology of magnetic fields in HAeBe stars has led to the conclusion that these objects are the pre-main sequence progenitors of main sequence Ap and Bp stars.  Ap and Bp stars are main sequence chemically peculiar stars with strong magnetic fields.  Their magnetic strengths range from 300 G to $\sim$10 kG (\cite[Auri\`ere et al. 2007]{Aurier2007}), and they have an incidence of 5-10\%.  The strong chemical peculiarities in these stars are understood to be a consequence of atomic diffusion operating in the stable radiative stellar envelope.  However, the timescale for the development of these peculiarities remains unknown.  Measurements of chemical abundances in the progenitors of Ap and Bp stars, magnetic HAeBe stars, would provide important constraints on this timescale.  However, the chemical abundances of HAeBe stars in general are poorly studied (e.g. \cite[Acke \& Waelkens 2004]{Acke2004}).  Therefore, we performed a precise investigation of chemical abundances in both magnetic and non-magnetic HAeBe stars.  

The $\lambda$ Bootis stars are chemically peculiar A-type stars, but unlike Ap stars they are characterized by underabundances of iron-peak elements and approximately normal abundances for C, N, O, and often S.  These stars are much rarer than Ap stars, appearing with roughly a 2\% incidence on the main sequence (\cite[Paunzen 2001]{Paunzen2001}).  The origin of these abundances is unknown, but they are unlikely to be a result of atomic diffusion (\cite[e.g. Charbonneau 1993]{Charbonneau1993}).  A leading hypothesis is that $\lambda$ Boo peculiarities are the result of selective accretion, in which lighter elements are accreted more readily than heavier elements, building up a layer of apparent underabundances at the surface of the star (\cite[Venn \& Lambert 1990]{Venn1990}).  
If the formation of $\lambda$ Boo peculiarities depends on an accretion process, then naively one would expect to find these peculiarities more often in HAeBe stars, objects which were recently accreting, and may still be.  Thus a second goal of this study is to search for the presence of $\lambda$ Boo peculiarities among HAeBe stars.  

Here we present results from \cite{Folsom2012}, and from \cite{Folsom2008}, providing a sample of 23 HAeBe stars, 4 of which have confirmed magnetic fields.  

\section{Observations}

To investigate these questions we used observations obtained with the ESPaDOnS instrument at the Canada-France-Hawaii Telescope.  This is a high resolution (R=65000) spectropolarimeter with a wavelength range of 3700-10500 \AA.  Observations for one southern target (HD~101412) were obtained with the SEMLPOL polarimeter attached to the University College London Echelle Spectrograph at the Anglo-Australian Telescope.  For the analysis of Balmer lines, archival observations from the FORS1 spectrograph at the Very Large Telescope were used.  This lower resolution instrument obtains spectra in a single order, which removes some ambiguities in the normalization of Balmer lines.  

The magnetic properties of the observations used here were analyzed by \cite{Alecian2013}, thus we have precise self-consistent diagnostics of presence or absence of magnetic fields in these stars.  The sample was chosen to include 23 stars, covering a range of 7500 to 15000 K in $T_{\rm eff}$, and spanning a range of $v \sin i$ up to 200 km\,s$^{-1}$.  The stars were also chosen to have only modest amounts of emission, or shell absorption, in their spectra.  This bias was necessary, as we required a large number of uncontaminated photospheric lines in order to determine photospheric chemical abundances.  The sample includes all well established magnetic HAeBe stars cooler than 15000 K.

\section{Abundance analysis}

The abundance analysis proceeded by directly fitting synthetic spectra, produced with the ZEEMAN spectrum synthesis code (\cite[Landstreet 1988]{Landstreet1988}, \cite[Wade et al. 2001]{Wade2001}), to the observed spectra.  Initial estimates of $T_{\rm eff}$ and $\log g$ were made by fitting the wings of Balmer lines, far from contamination by emission.  However, there is a substantial degeneracy between  $T_{\rm eff}$ and $\log g$ in these estimates, thus they were refined using ionization and excitation balances, by simultaneously fitting lines of an element with different ionization states and a wide range of excitation potentials.  Chemical abundances, $v \sin i$, microturbulence, $T_{\rm eff}$, and $\log g$ were fit simultaneously by $\chi^2$ minimization.  Six spectral regions, $\sim$500 \AA\ long, were independently fit for each star.  The average and standard deviation of these results were taken as the final best value and its uncertainty, respectively.  

Great care was taken to avoid fitting lines contaminated with circumstellar emission or absorption.  With multiple observations of the stars, we could often identify lines contaminated by small amounts of emission using unexpected line variability.  By comparing line shapes to the synthetic spectra, we could identify lines that departed from simple rotation broadening, and exclude those.  Finally, by examining lines with very low excitation potentials for inconsistencies, we could identify potential emission infilling and exclude those lines from the fit.  

For further details on the analysis methodology see \cite{Folsom2012}.  

\section{Results}

From the abundance analysis we find 10 (out of 23) stars are chemically normal, with approximately solar abundances.  However, we find another 11 stars that show underabundances of iron-peak elements and roughly solar abundances of C, N, and O, thus they display $\lambda$ Boo peculiarities with varying strengths.  This represents roughly a 50\% incidence of $\lambda$ Boo peculiarities, while these peculiarities only appear in about 2\% of main sequence A stars (\cite[Paunzen 2001]{Paunzen2001}).  

Among the four magnetic stars in our sample, we find one that is chemically normal (HD~190073), one that displays  $\lambda$ Boo peculiarities (HD~101412), one with weak marginal Bp peculiarities (V380~Ori~A), and one with strong Bp peculiarities (HD~72106~A).  The presence of  $\lambda$ Boo peculiarities in HD 101412 appears to be simply a consequence of the very high incidence of these peculiarities among HAeBe stars, thus we consider both HD~190073 and HD~101412 to be chemically indistinguishable from the non-magnetic stars.  Examining the binary system HD~72106 in detail, we find all the Herbig star characteristics of the system are associated with the secondary. Placing the system on the H-R diagram, the primary is consistent with the zero age main sequence (ZAMS).  Thus we conclude that HD~72106~A has probably reached the main sequence, but it is likely still a very young object since the secondary still appears to be a pre-main sequence star.  For the SB2 system V380~Ori, both components are clearly still on the pre-main sequence.  In both binary systems with a magnetic primary, we find abundances for the secondary consistent with solar.  

The full set of final atmospheric parameters and chemical abundances derived for all stars are presented by \cite{Folsom2012}, and by \cite{Folsom2008} for HD~72106.

\begin{figure}[bht]
\begin{center}
 \includegraphics[width=3.0in]{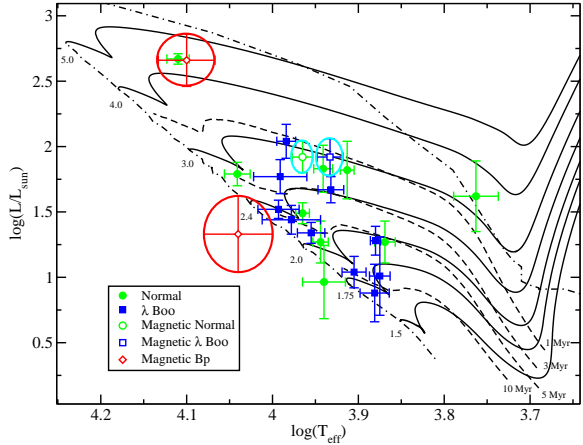} 
 \caption{H-R diagram for the stars in this study.  Solid lines are evolutionary tracks (labeled by mass in $M_{\odot}$), dashed lines are isochrones (labeled by age), and the birthline and ZAMS are dashed-dotted lines.  The stars are classified by their chemical abundances and magnetic properties, and the four magnetic stars are highlighted with ovals.   }
   \label{fig-hrd}
\end{center}
\end{figure}

\section{Discussion and Conclusions}

We find roughly 50\% of the stars in our sample display $\lambda$ Boo chemical peculiarities, an incidence rate dramatically larger than the $\sim$2\% seen on the main sequence.  We interpret this as evidence in favor of a selective accretion hypothesis for the formation of $\lambda$ Boo peculiarities.  If the mechanism for forming $\lambda$ Boo peculiarities depends on accretion, then one would expect to find such peculiarities frequently in HAeBe stars, objects which have recently been accreting, and may still be.  Over time these surface chemical peculiarities would become mixed into the stars, and thus on the main sequence the stars would likely display normal abundances.  

Among the magnetic stars, we find two stars with Bp chemical peculiarities and two stars with abundances matching the non-magnetic HAeBe stars (i.e. chemically `normal').  Thus Bp peculiarities can appear on the pre-main sequence, but magnetic fields precede the presence of Ap/Bp peculiarities.  This is in contrast to the main sequence where all strongly magnetic A and late B stars are Ap and Bp stars.  Placing the magnetic stars on the H-R diagram, there appears to be a rough progression, with the two chemically `normal' stars being further from the ZAMS, the weak Bp star being closer to the ZAMS, and the strong Bp star being on the ZAMS.  This may represent the development of chemical peculiarities over time during the pre-main sequence, however a larger sample size is needed to draw firm conclusions.  

These results agree well with the work of \cite{Cowley2010} who independently found $\lambda$ Boo peculiarities in HD~101412, and \cite{Cowley2012} who found solar abundances in HD~190073.  \cite{Cowley2013} and Cowley et al. (in press) report an additional chemically normal HAeBe star and a $\lambda$ Boo HAeBe star, further supporting the very high incidence of $\lambda$ Boo peculiarities among HAeBe stars.  

Finally, we find no evidence for other types of chemical peculiarities among the HAeBe stars in our sample.  Specifically, we see no Am or HgMn chemical peculiarities.  These chemical peculiarities occur in 10 to 20\% of main sequence A- and late B-type stars, and thus naively we would expect to have found between 2 and 4 such stars.  A much larger sample is needed to draw any firm conclusions, but it may be that these peculiarities do not have time to form on the pre-main sequence.

\end{document}